# An iron(III) phosphonate cluster containing a nonanuclear ring

Hong-Chang Yao,[†] Jun-Jie Wang,[†] Yun-Sheng Ma,[†] Oliver Waldmann,[‡,*] Wen-Xin Du,[¶] You Song,[†] Yi-Zhi Li,[†] Li-Min Zheng[†,*] Silvio Decurtins,[‡,*] Xin-Quan Xin[†]

[†] *State Key Laboratory of Coordination Chemistry, Coordination Chemistry Institute, Nanjing University, Nanjing 210093, P. R. China*
[‡] *Department of Chemistry and Biochemistry, University of Bern, Freiestrasse 3, CH-3012 Bern, Switzerland*
[¶] *State Key Laboratory of Structure Chemistry, Fujian Institute of Research on the Structure of Matter, The Chinese Academy of Sciences, Fuzhou 350002, P. R. China*

**This paper reports the first example of cyclic ferric clusters with an odd number of iron atoms capped by phosphonate ligands, namely, [Fe$_9$($\mu$-OH)$_7$($\mu$-O)$_2$(O$_3$PC$_6$H$_9$)$_8$(py)$_{12}$]. The magnetic studies support a S = 1/2 ground state with an exchange coupling constant of about J ≈ -30 K.**

Cyclic iron cages are attractive not only because of their capability to accommodate alkali-metal cations[1,2] but also because of their unprecedented magnetic behavior.[3] Having been regarded initially as ideal models for studying 1D anti-ferromagnetic materials it subsequently became clear that their magnetism is best described by a quantized rotation of the Néel-vector and is more similar to that of 2D and 3D materials.[4] Moreover, quantum phenomena such as Néel-vector tunneling were also demonstrated.[5] However, all cyclic iron cages recorded so far are even-membered wheels with 6,[1,2] 8,[2,6] 10,[3,7] 12[8] and 16[9] metal ions in which carboxylate ligands are usually involved in linking the metal ions. Synthesizing odd-membered iron rings is an interesting goal, not only from a chemical perspective but also because the odd number of magnetic electrons induce quantum spin-frustration effects.[10] Recently, a nona-nuclear wheel consisting of one Ni(II) and eight Cr(III) ions has been synthesized.[11] Because of the even number of electrons, the frustration situation in such a wheel is fundamentally different from that in systems with an odd number of electrons.[12] In recent years iron clusters involving phosphonate ligands have attracted much attention. Several complexes of iron phosphonates with Fe$_4$, Fe$_6$, Fe$_7$ and Fe$_{14}$ cages have been obtained.[13,14] The oxo-centered [Fe$_3$($\mu_3$-O)]$^{7+}$ triangle units are found in these cages and the carboxylates appear as coligands. Herein, we report a novel nonanuclear iron(III) cage complex, [Fe$_9$($\mu$-OH)$_7$($\mu$-O)$_2$(O$_3$PC$_6$H$_9$)$_8$(py)$_{12}$]·6H$_2$O (**1**·6H$_2$O), which contains a nine iron ring of {Fe$_9$($\mu$-OH)$_7$($\mu$-O)$_2$} capped by phosphonate ligands.

Treatment [Fe$_3$O(O$_2$CMe)$_6$(H$_2$O)$_3$]NO$_3$ (0.262 g, 0.4 mmol) with two equivalents of C$_6$H$_9$PO$_3$H$_2$ (0.130 g, 0.8 mmol) in pyridine (10 mL) gave a light brown solution from which green-yellow block crystals were formed in good yield (46%) by diffusing Et$_2$O into the filtrate.† The purity was judged by comparing powder XRD with the crystallographic result. The $^{57}$Fe Mössbauer spectrum of **1** was recorded in zero field at room temperature. The observed broad doublet can be best fitted to two doublets with an area ratio of 2:1 with isomer shifts ($\delta$) and quadrupole splitting ($\Delta E_Q$) of 0.359, 0.579 mm/s and 0.447, 0.578 mm/s, respectively, corresponding to two types of high-spin ferric centers.[15]

Single-crystal structural analysis reveals that complex **1**·6H$_2$O‡ crystallizes in the monoclinic space group *C*2. The asymmetric unit contains half of [Fe$_9$($\mu$-OH)$_7$($\mu$-O)$_2$(O$_3$PC$_6$H$_9$)$_8$(py)$_{12}$]. The lattice water molecules reside on seven crystallographic sites with partial occupancies which are summed to "three". There are five crystallographically different Fe atoms in the structure. The Fe(1) and O(5) atoms lie on a twofold axis; the other four Fe atoms are on general positions. Two types of Fe atoms, A and B, can be distinguished, see Figure 1 (top). In type A [Fe(2), Fe(3), Fe(5)], the distorted octahedral sphere of each Fe atom is completed by three phosphonate oxygen, two $\mu$-OH (or $\mu$-O) oxygen and one pyridine nitrogen atom, forming a FeO$_5$N coordination environment. In type B [Fe(1), Fe(4)], the octahedral sphere of each Fe is completed by two phosphonate oxygen, two $\mu$-OH (or $\mu$-O) oxygen and two pyridine nitrogen atoms, forming coordination mode of FeO$_4$N$_2$. The molar ratio of type A and B Fe atoms is 2:1, in agreement with the Mössbauer result. The Fe–O bond lengths are in the range of 1.950(4) –2.098(4) Å. The Fe-N bond distances are between 2.192(5) and 2.248(6) Å.

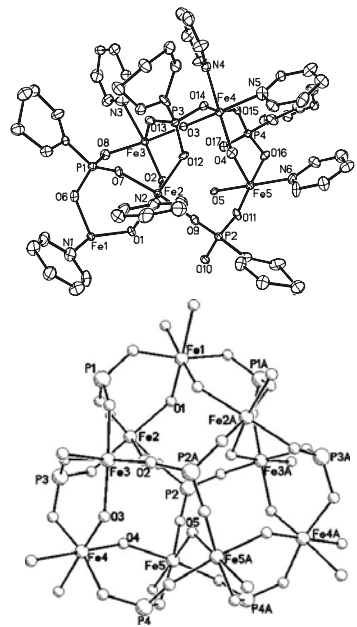

**Fig. 1**. Top: Asymmetric unit of **1** (ellipsoids at 30% probability); Bottom: Inorganic core of **1**. Hydrogen atoms are omitted.

Adjacent iron atoms are connected through $\mu$–OH or $\mu$–O ligands, forming a twisted 18-member ring of {Fe$_9$($\mu$-OH)$_7$($\mu$-O)$_2$}. Within this ring, the Fe–O bond lengths are in the range 1.965(5) – 2.098(4) Å and the Fe–O–Fe bond angles vary between 119.8(2) – 137.2(2)º (Fig. 1). The ring is capped by two equivalent phosphonate groups [P(2)] through the coordination of the phosphonate oxygen atoms [O(9), O(10A), O(11)] to the iron atoms Fe(2), Fe(3A) and Fe(5). Three neighboring iron atoms, e.g., Fe(1)/Fe(2)/Fe(3), Fe(2)/Fe(3)/Fe(4), and Fe(4)/Fe(5)/Fe(5A), are each capped by one phosphonate ligand, forming a neutral inorganic cage of [$Fe_9(\mu\text{-}OH)_7(\mu\text{-}O)_2(\mu\text{-}O_3P)_8$]. The cage is confined within a lipophilic shell composed of either phenyl groups protruded from phosphorus or pyridine.

\* lmzheng@nju.edu.cn, oliver.waldmann@iac.unibe.ch, silvio.decurtins@iac.unibe.ch

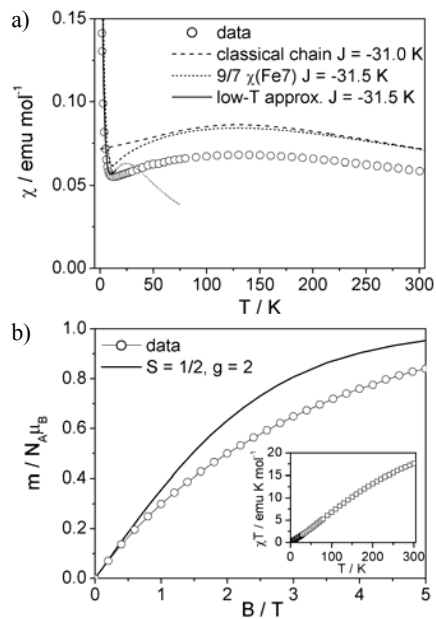

**Fig. 2**. (a) Magnetic susceptibility of **1**·6$H_2O$ vs. temperature (open circles). Lines represent results for the classical chain, $Fe_7$ ring, and low-temperature approximation discussed in the text. (b) Magnetic moment of **1**·6$H_2O$ vs. field at 1.8 K (open circles). The solid line represents the expectation for a S = 1/2 level. The inset presents $\chi T$ vs. temperature.

The magnetic susceptibility was measured in the 1.8 – 300 K temperature range.§ The $\chi(T)$ curve exhibits a maximum at ca. 130 K, and a strong increase at the lowest temperatures (Fig. 2a). The maximum and the decrease of $\chi T$ with temperature (inset of Fig. 2b) clearly indicate significant anti-ferromagnetic interactions between the metal ions within a cluster. The $\chi T$ value at 300 K (17.6 emu K mol$^{-1}$) is much lower than the spin-only value of 39.37 emu K mol$^{-1}$ for nine non-interacting high-spin Fe(III) ions (S = 5/2, g = 2.0). On cooling, $\chi T$ approaches a value of ca. 0.30 emu K mol$^{-1}$, suggesting a S = 1/2 ground state (0.375 emu K mol$^{-1}$). A S = 1/2 cluster ground state is clearly confirmed by the magnetization curve recorded at 1.8 K (Fig. 2b).

The anti-ferromagnetic exchange interactions in **1** are mainly mediated through two pathways, namely through the $\mu$-OH or $\mu$-O bridges within the {$Fe_9(\mu\text{-}OH)_7(\mu\text{-}O)_2$} ring and the O-P-O units which cap neighboring three Fe atoms. It is well known that in comparison to oxide or hydroxide bridges three-atom bridges such as O-P-O only support negligible exchange interactions.[14,16,17] The $\mu$-OH or $\mu$-O pathways are thus expected to be dominant in **1**. The Fe-Fe distances across the $\mu$-OH (3.695 – 3.757 Å) and $\mu$-O (3.529 – 3.532 Å) bridges are quite close and the *J* value is not very sensitive to the Fe–O–Fe angle in the range of 120 – 180°.[16,18]

The analysis of the magnetism in **1** is very challenging because of the huge dimension of the Hilbert space (10,077,696), which prevents an exact calculation of the magnetic susceptibility. Qualitatively, however, the situation is as follows. For an odd ring of half-integer spin centers, the ground state is four-fold degenerate (or quasi degenerate), forming a doublet of two S = 1/2 levels.[12] The reason for the S = 1/2 ground state is easily explained: For an odd number of magnetic electrons the spin-coupling rules allow only half-integer values of the total spin S. The anti-ferromagnetic interactions select the lowest possible value of S, hence S = 1/2 (for a ring with even number of electrons the total spin assumes integer values and the ground state is thus S = 0). Accordingly, the susceptibility exhibits the broad maximum typical for anti-ferromagnetic wheels, but an additional Curie behavior at low temperatures corresponding to a S = 1/2 ground state. These characteristic features of an odd half-integer-spin ring are clearly present in **1**, confirming its odd-membered cyclic structure.

In view of the structure, **1** may, in a first approximation, be described by a spin-5/2 ring with a $C_9$ symmetry of the next-neighbor Heisenberg exchange interactions,

$$H = -J\left(\sum_{i=1}^{8} \mathbf{S}_i \cdot \mathbf{S}_{i+1} + \mathbf{S}_9 \cdot \mathbf{S}_1\right)$$

For the quantitative analysis, one has to resort to approximate techniques for evaluating the susceptibility; three such approximations will be discussed. Their applicability and/or limitations were inferred from investigations on nine-membered spin-3/2 and spin-1/2 rings and seven-membered spin-5/2, spin-3/2, and spin-1/2 rings, for which the susceptibility could be calculated exactly. In a first approach, the spin-5/2 centers are replaced by classical spins. The susceptibility of a classical $Fe_9$ ring can be calculated analytically,[19] but we found that a better approximation is given by the well-known classical chain formula.[20] The result for J = -31.0 K (g = 2.0 will be always assumed) is shown in Fig. 2a. The classical curve approaches a finite value at zero temperature, which is a well-known artifact of the classical approximation, and hence is of no significance. Another approximation to the susceptibility of a $Fe_9$ ring is obtained by multiplying the susceptibility of a $Fe_7$ ring, denoted as $\chi_{Fe_7}(T)$, by a factor of 9/7 in order to correct for the different number of spin centers. Our calculations on the smaller rings showed that the resulting curve should be a reasonable approximation to the exact susceptibility of a $Fe_9$ ring. The curve for J = -31.5 K is presented in Fig. 2a. Finally, the susceptibility at low-temperatures was calculated by using sparse-matrix diagonalization techniques. The result for J = -31.5 K is accurate to within 1% for temperatures up to 15 K, but strongly deviates at the higher temperatures (Fig. 2a).

The comparison of the approximate results with the experimental $\chi(T)$ curve is clear in its conclusions. The maximum, the Curie-tail at the lowest temperatures, and the overall behavior is well described. Only the absolute value of the susceptibility is too low. This is attributed to the fact, that in reality **1** does not exhibit a $C_9$ symmetry as assumed above. The structure of the cluster permits a variation of the coupling constants along the ring. The approximate three-fold symmetry suggests that every third coupling has the strength J', and that the Hamiltonian

$$H' = H - (J' - J)(\mathbf{S}_2 \cdot \mathbf{S}_3 + \mathbf{S}_5 \cdot \mathbf{S}_6 + \mathbf{S}_8 \cdot \mathbf{S}_9)$$

should be a more realistic approach. Unfortunately, the classical chain formula and the $Fe_7$ approximation cannot be applied here. Thus, the susceptibility of a nine-membered spin-3/2 ring was numerically investigated. Figure 3 presents the theoretical susceptibility for J'/J = 1, 1.5, and 2, showing that a modulation as small as a factor of two easily produces a lowering of χ(T) as it is observed experimentally. We also checked models taking into account the actual two-fold symmetry of the cluster and exchange paths corresponding to the O-P-O units. Within reasonable limits, no significant effects were obtained. It is thus concluded that the exchange coupling in **1** is about J ≈ -30 K, and that a small variation in the coupling constants consistent with the approximate three-fold symmetry is present. Compared with the coupling constants for other Fe(III) systems linked by μ-O (J ca. -100 cm$^{-1}$) or μ-OH (J ca. -10 cm$^{-1}$) bridges,[21] the estimated J value is reasonable and is close to those found in other ferric wheels.[8]

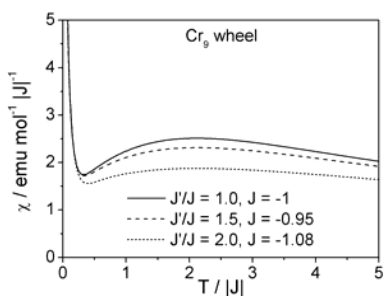

**Fig. 3**. Theoretical χ(T) curves for a nine-membered spin-3/2 ring in units of J (see Hamiltonian H′). The results for three ratios J'/J are displayed. The value of J has been slightly scaled in order to keep the maximum at the same temperature.

In summary, this paper reports the structure and magnetic properties of a new nonanuclear iron(III) cluster [Fe$_9$(μ-OH)$_7$(μ-O)$_2$(O$_3$PC$_6$H$_9$)$_8$(py)$_{12}$]. The significance of the compound is two fold. Firstly, it provides the first example of a cyclic ferric cluster with an odd number of iron atoms. Because of the odd number of electrons, an unprecedented situation concerning quantum spin-frustration is realized in this wheel. And secondly, to our knowledge, it is the first example of iron phosphonate clusters without carboxylate as coligands. Further work is in progress to search for new iron phosphonate clusters with new topologies and interesting magnetic properties.

**Acknowledgement.** We thank the NNSF of China (No. 20325103), the Ministry of Education of China and the NSF of Jiangsu province (No. BK2002078) for financial supports, and Dr. H.-B. Huang for the Mössbauer measurement.

## Notes and references

† Elemental analysis calcd(%) for **1**·6H$_2$O (C$_{108}$H$_{151}$N$_{12}$O$_{39}$P$_8$Fe$_9$): C, 43.36; H, 5.09; N, 5.62. Found: C, 43.75; H: 5.18; N: 5.96.

‡Crystal Data for **1**·6H$_2$O: monoclinic, space group *C*2, *a* = 26.268 (1), *b* = 18.929 (1), *c* = 16.873(1) Å, *β* = 118.462 (1) °, *V* = 7375.7(3) Å$^3$, *Z* = 2, *Mr* = 2991.8, *ρ* = 1.347 g cm$^{-1}$, *F*(000) = 3098, *μ*$_{MoKα}$ = 1.018 mm$^{-1}$. The data collection was carried on a Bruker SMART 1K CCD diffractometer using graphite-monochromatized Mo*Kα* radiation (λ = 0.71073 Å). The structure was solved by direct methods and refined on *F*$^2$ by full-matrix least squares using SHELXTL,[22] and converging at R1 = 0.0578, *w*R2 = 0.1151.

§Temperature and field dependent magnetic data were recorded with a SQUID magnetometer (Quantum Design). The susceptibility measurements were done at 2000 G. A number of powder, polycrystalline, and single-crystal samples were investigated in order to confirm that the reported values of χ(T) are not affected by the typical problems, such as loss of solvent, degradation, impurities, etc.

1. A. Caneschi, A. Cornia, S. J. Lippard, *Angew. Chem. Int. Ed. Engl.* 1995, 34, 467-469.
2. R. W. Saalfrank, I. Bernt, E. Uller, F. Hampel, *Angew. Chem. Int. Ed. Engl.* 1997, 36, 2482.
3. K. L. Taft, C. D. Delfs, G. C. Papaefthymiou, S. Foner, D. Gatteschi, S. J. Lippard, *J. Am. Chem. Soc.* 1994, 116, 823-832.
4. O. Waldmann, *Phys. Rev. B* 2001, 65, 024424; O. Waldmann, T. Guidi, S. Carretta, C. Mondelli, A. L. Dearden, *Phys. Rev. Lett.* 2003, 91, 237202. O. Waldmann, *Coord. Chem. Rev.* 2005, 249, 2550.
5. O. Waldmann, C. Dobe, H. Mutka, A. Furrer, H. U. Güdel, *Phys. Rev. Lett.* 2005, 95, 057202.
6. C. Cañada-Vilalta, T. A. O'Brien, M. Pink, E. R. Davidson, G. Christou, *Inorg. Chem.* 2003, 42, 7819-7829; C. Cañada-Vilalta, M. Pink, G. Christou, *Chem. Commun.* 2003, 1240-1241.
7. C. Benelli, S. Parson, G. A. Solan, R. E. P. Winpenny, *Angew. Chem. Int. Ed. Engl.* 1996, 35, 1825- 1828; K. L. Taft, S. J. Lippard, *J. Am. Chem. Soc.* 1990, 112, 9629-9630; S.-X. Liu, S. Lin, B.-Z. Lin, C.-C. Lin, J.-Q. Huang, *Angew. Chem. Int. Ed. Engl.* 2001, 40, 1084-1087.
8. A.-A. H. Abu-Nawwas, J. Cano, P. Christian, T. Mallah, G. Rajaraman, S. J. Teat, R. E. P. Winpenny, Y. Yukawa, *Chem. Commun.* 2004, 314-315; C. P. Raptopoulou, V. Tangoulis, E. Devlin, *Angew. Chem. Int. Ed. Engl.* 2002, 41, 2386-2389; G. L. Abbati, A. Caneschi, A. Cornia, A. C. Fabretti, D. Gatteschi, *Inorg. Chim. Acta*, 2000, 297, 291-300.
9. L. F. Jones, A. Batsanov, E. K. Brechin, D. Collison, M. Helliwell, T. Mallah, E. J. L. McInnes, S. Piligkos, *Angew. Chem. Int. Ed. Engl.* 2002, 41, 4318-4321.
10. D. Gatteschi, R. Sessoli, A. Cornia, *Chem. Commun.* 2000, 725-732.
11. O. Cador, D. Gatteschi, R. Sessoli, F. K. LArsen, J. Overgaard, A.-L. Barra, S. J. Teat, G. A. Timco, R. E. P. Winpenny, *Angew. Chem. Int. Ed. Engl.* 2004, 43, 5196.
12. J. C. Bonner, M. E. Fisher, *Phys. Rev.* 1964, 135, A640; K. Bärwinkel, P. Hage, H.-J. Schmidt, J. Schnack, *Phys. Rev. B* 2003, 68, 054422.
13. E. I. Tolis, M. Helliwell, S. Langley, J. Raftery, R. E. P. Winpenny, *Angew. Chem. Int. Ed. Engl.* 2003, 42, 3804-3808.
14. H.-C. Yao, Y.-Z. Li, L.-M. Zheng, X.-Q. Xin, *Inorg. Chim. Acta*, 2005, 358, 2523-2529.
15. B. P. Murch, F. C. Bradley, P. D. Boyle, V. Papaefthymiou, L. Que, Jr., *J. Am. Chem. Soc.* 1987, 109, 7993.
16. R. Werner, S. Ostrovsky, K. Griesar, W. Haase, *Inorg. Chim. Acta*, 2001, 326, 78-88.
17. R. E. Normam, S. Yan, L. Que, G. Baches, J. Ling, J. Sanders-Loehr, J. H. Zhang, C. J. O'Connor, *J. Am. Chem. Soc.* 1990, 112, 1554.
18. R. E. Normam, R. C. Holz, S. Menage, C. J. O'Connor, J. H. Zhang, L. Que, Jr., *Inorg. Chem.*, 1990, 29, 4629-4637; R. Hotzelman, K. Wieghard, U. Flörke, H.-J. Haupt, D. C. Weatherburn, J. Bonvoisin, G. Blondin, J.-J. Girerd, *J. Am. Chem. Soc.* 1992, 114, 1681-1696.
19. G. S. Joyce, *Phys. Rev.* 1967, 155, 478.
20. M. E. Fisher, *Am. J. Phys.* 1964, 32, 343; O. Kahn, *Molecular Magnetism*, VCH Publishers, Inc., New York, 1993.
21. S. M. Gorun, S. J. Lippard, *Inorg. Chem.* 1991, 30, 1625-1630.
22. "SHELXTL (version 5.0) Reference Manual." Siemens Industrial Automation, Analytial Instrumentation, Madison, WI, 1995.